\documentclass[preprint,showpacs,preprintnumbers,amsmath,tightenlines,prb]{revtex4}

\usepackage{graphicx}
\usepackage{dcolumn}
\usepackage{bm}

\begin{document}


\title{High order Chin actions in path integral Monte Carlo}

\author{K. Sakkos}
\author{J. Casulleras}%
\author{J. Boronat}
\affiliation{
Departament de F\'{\i}sica i Enginyeria Nuclear,\\
Universitat Polit{\`e}cnica de Catalunya,\\
Campus Nord B4--B5, E--08034 Barcelona, Spain
}%

\date{\today}

\begin{abstract}
High order actions proposed by Chin have been used for the first time in path
integral Monte Carlo simulations. Contrarily to the Takahashi-Imada action,
which is accurate to fourth order only for the trace, the Chin action is fully
fourth order, with the additional advantage that the leading fourth and sixth
order error coefficients are finely tunable. By optimizing two free parameters
entering in the new action we show that the time step error dependence achieved
is best fitted with a sixth order law. The computational effort per bead
is increased but the total number of beads is greatly reduced, and the
efficiency improvement with respect to the primitive approximation is
approximately a factor of ten. The Chin action is tested in a one-dimensional
harmonic oscillator, a H$_2$ drop, and bulk liquid $^4$He. In all cases a
sixth-order law is obtained with values of the number of beads that compare
well with the pair action  approximation in the stringent test of superfluid
$^4$He.   \end{abstract}

\maketitle

\section{\label{sec:introduction}Introduction}

Path integral Monte Carlo (PIMC) methods provide a fundamental
approach in the study of interacting quantum many-body systems 
at low temperatures.~\cite{gillan,chakra,ceperley} This accurate
simulation tool relies on the well-known convolution property of the
thermal density matrix which allows for estimating the density matrix
at low temperature from their knowledge at higher temperature, where the
system is well described by classical statistical
mechanics.~\cite{fey1,fey2,kleinert} The partition
function $Z$ of the quantum system is then written as a multidimensional
integral with a distribution law that resembles the one of a closed
classical polymer with an inter-bead harmonic coupling. 
If one either ignores the quantum statistics of the particles (\textit{boltzmanons}),
or they are bosons, then the statistical distribution law is positive definite and
can be interpreted as a probability distribution function which can
be sampled by standard Metropolis Monte Carlo methods. The mapping of this
finite-temperature quantum system to a classic system of polymers was already
suggested by Feynman~\cite{fey2} and implemented in practice by
Barker,~\cite{barker} and Chandler and Wolynes~\cite{chandler} in their
pioneering works.

In the simplest and most common implementations of the PIMC method the
density matrix at high temperature, which constitutes the building block of
the polymer chain, is considered fully classical, i.e., kinetic and
potential parts of the action are splitted ignoring any contribution
arising from the non-conmutativity of the kinetic $\hat{K}$ and potential
$\hat{V}$ operators. This approximation is known as primitive action
(PA) and is accurate to order $(\beta /M)^2$, with $\beta$ the inverse
temperature and $M$ the number of convolution terms (beads). When the temperature
decreases the number of beads to reach convergence in terms of $1/M$ increases 
and the system becomes more and more quantum. In the classical
limit, PIMC reduces to classic MC since the polymeric chain reduces to a
single point ($M=1$). On the other hand, near the zero-temperature limit 
the chains acquire extensions comparable to the size of the simulation box.
PA is accurate to study semiclassical systems in which the quantum
effects are comparatively small and therefore the number of beads in the
asymptotic regime is low enough for an efficient sampling. However, if the 
interest lies in a fully quantum regime at very cold temperatures, $M$
increases very fast making simulations hard, if not impossible, due to 
very low efficiency in the sampling of the long chains involved.  

The study of fully quantum fluids and solids require better actions than
the PA. An accurate action to deal with very low temperatures, down to
superfluid regimes, relies on the pair-product approximation~\cite{pollock}
 in which the
basic piece of the PIMC chain is the exact action for two isolated
particles. The pair-product action has been used extensively in the study
of superfluids and it is specially accurate for hard-spherelike systems
such as liquid $^4$He.~\cite{ceperley} However, its use for non-radial interactions is much
more difficult due to the complexity of the pair density matrix. A
different approach to improve the PA relies on the use of higher-order actions
obtained directly from the exponential of the Hamiltonian, $\exp(-\beta
\hat{H})$. Working in this direction, Takahashi and Imada~\cite{takahashi} and later on, and
independently, Li and Broughton~\cite{li} managed to write a new action (TIA) which
is accurate to fourth order $(\beta/M)^4$ for the trace.
To this end, they incorporated in the expression for the action the double
commutator $[[\hat{V},\hat{K}],\hat{V}]$.  More recently, Jang
\textit{et al.}~\cite{jang} proposed a new action based on the Suzuki
factorization~\cite{suzuki} and
accurate to fourth order. In spite of being a real fourth order action, and
therefore better than the TIA, the results obtained by Jang \textit{et
al.}~\cite{jang}
do not show a significant improvement towards the reduction of the number
of beads in the asymptote $\beta / M \rightarrow 0$ respect to the TIA.     

In the present work, we introduce a new family of high order actions based
on the symplectic developments of Chin~\cite{chin} and Chin and
Chen~\cite{chin2} that have already
proved its efficiency in solving the Schr\"odinger equation,~\cite{chin3} problems in 
classical mechanics,~\cite{chin4} and in the implementation of evolution operators in
density functional theory.~\cite{kro} In Chin's factorization of the evolution
operator all the coefficients are $i)$  positive (and therefore directly
implementable in Monte Carlo simulations) and $ii)$ continuously tunable, which
makes possible to force the error terms of fourth order to roughly
cancel each other for the specific range of time-step values of interest for the
actual simulations. We have introduced this new
action in our PIMC algorithm and all the results  presented in this article
show that the accuracy achieved in terms of $\beta/M$ is sixth order in
practice in spite of the fact that the action is only accurate to fourth
order. The attained reduction in the number of beads makes this algorithm
competitive with the pair-product approximation in the study of quantum
fluids and solids at ultralow temperatures.

The rest of the work is organized as follows. The action and the method
used for the PIMC simulation is contained in Sec. II. In Sec. III, we
present results obtained with the Chin's action for the one-dimensional
harmonic oscillator, a drop of H$_2$ molecules, and bulk liquid $^4$He at
low temperatures. Finally, a brief summary and the main conclusions are
reported in Sec. IV.

\section{\label{sec:method} Formalism}

At finite temperature, the knowledge of the quantum partition function
\begin{equation}
Z = \text{Tr} \, e^{-\beta \hat{H}}
\label{zeta}
\end{equation}
allows for a full microscopic description of the properties of a given
system, with $\beta=1/T$ and $\hat{H} = \hat{K} + \hat{V}$ the Hamiltonian.
In a $N$-particle system, the kinetic operator is given by
\begin{equation}
\hat{K} = -\frac{\hbar^2}{2m} \sum_{i=1}^{N} \bm{\nabla}_i^2 \ ,
\label{kinetic}
\end{equation} 
and the potential one, assuming pairwise interactions, by
\begin{equation}
\hat{V} = \sum_{i<j}^{N} V(r_{ij}) \ .
\label{potential}
\end{equation}
The non-conmutativity of the quantum operators $\hat{K}$ and $\hat{V}$ makes
impractical a direct calculation of the partition function $Z$
(\ref{zeta}). Instead, all practical implementations intended for Monte
Carlo estimations of $Z$ rely on its convolution approximation
\begin{equation}
e^{-\beta (\hat{K} + \hat{V})} = \left( e^{-\varepsilon ( \hat{K} + \hat{V}
)} \right)^M \ ,
\label{convolution}
\end{equation}
with $\varepsilon = \beta/M$, where  each one of the terms in the r.h.s.
corresponds to a higher temperature $M T$. In the most simple
approximation, known as primitive action (PA), the kinetic and potential
contributions factorize
\begin{equation}
e^{-\varepsilon ( \hat{K} + \hat{V})} \simeq e^{-\varepsilon  \hat{K}} \,
e^{-\varepsilon  \hat{V}} \ ,
\label{primitive}
\end{equation}
the convergence to the exact result being warranted by the Trotter
formula~\cite{trotter}
\begin{equation}
e^{-\beta (\hat{K} + \hat{V})}  = \lim_{M \to \infty} \left(
e^{-\varepsilon  \hat{K}}  \, e^{-\varepsilon  \hat{V}}  \right)^M \ .
\label{trotter}
\end{equation} 

The primitive action is a particular case of a more general form in which one
can decompose the action, that is
\begin{equation}
e^{-\varepsilon (\hat{K} + \hat{V})} \simeq \prod_{i=1}^{n} e^{-t_i
\varepsilon \hat{K}} \, e^{-v_i \varepsilon \hat{V}} \ ,
\label{suzukin}
\end{equation}
with $\{ t_i,v_i \}$ parameters to be determined according to the required
accuracy of the approximation. As we are interested in the Monte Carlo
implementation of Eq. (\ref{suzukin}), all these parameters must be positive.
However, the Sheng-Suzuki~\cite{sheng,suzuki2} theorem proves that this is impossible beyond
second order in $\varepsilon$. Recently, Chin~\cite{chin} has analyzed these expansions
(\ref{suzukin}) for the estimation of the quantum partition function where
only the trace is required (\ref{zeta}). As the trace is invariant under a
similarity transformation $S$, it could be possible to improve the order of the
approximation by a proper choice of $S$. However, Chin~\cite{chin} has proved that
expansions like (\ref{suzukin}) can not be corrected with $S$ beyond second
order, generalizing in this form the Sheng-Suzuki
theorem.~\cite{sheng,suzuki2}

In order to overcome the limitation imposed by the
Sheng-Suzuki~\cite{sheng,suzuki2} theorem it
is necessary to include in the operator expansion terms with double
commutators such as $[[\hat{V},\hat{K}],\hat{V}]$. In a recent work,
Chin~\cite{chin}
has proved that a second-order algorithm (PA) can be corrected by a
similarity transformation if that commutator is introduced. The result  
yields the TIA, with a trace that is accurate 
to fourth order. The TIA improves significantly the accuracy of the PA in
PIMC simulations but not enough to deal properly with fully quantum fluids.
In order to make a step further it is necessary to work directly with
real fourth-order actions. Chin and Chen~\cite{chin2}  have worked out a continuous family of gradient
symplectic  algorithms which are accurate to fourth order and that have
proved to be extremely accurate in the resolution of classical and quantum
problems. Now, we extend the focus of its applicability to the PIMC
algorithm.

The fourth-order action we have used is a two-parameter model given
explicitely by~\cite{chin2}
\begin{equation}
e^{-\varepsilon \hat{H}} \simeq e^{- v_1 \varepsilon \hat{W}_{a_1}}
e^{- t_1 \varepsilon \hat{K}} e^{- v_2 \varepsilon \hat{W}_{1-2 a_1}}
e^{- t_1 \varepsilon \hat{K}} e^{- v_1 \varepsilon \hat{W}_{a_1}}
e^{- 2 t_0 \varepsilon \hat{K}} \ ,
\label{chinaction}
\end{equation}   
hereafter referred as Chin action (CA). The generalized potentials
$\hat{W}$ resemble the one that appears in the TIA since both incorporate
the double commutator
\begin{equation}
[[\hat{V},\hat{K}],\hat{V}] = \frac{\hbar^2}{m} \, \sum_{i=1}^{N}
|\bm{F}_i|^2 \ ,
\label{commutator}
\end{equation}
with $\bm{F}_i$ the \textit{force} acting on particle $i$,
\begin{equation}
\bm{F}_i = \sum_{j \neq i}^{N} \bm{\nabla}_i V(r_{ij})  \ .
\label{force}
\end{equation}
The potentials $\hat{W}$ in Eq. (\ref{chinaction}) are explicitely
\begin{eqnarray}
\hat{W}_{a_1} & = & \hat{V} + \frac{u_0}{v_1} \, a_1 \varepsilon^2 \left( \frac{\hbar^2}{m} \, \sum_{i=1}^{N}
|\bm{F}_i|^2  \right) \label{doblev1} \\ 
\hat{W}_{1-2 a_1} & = & \hat{V} + \frac{u_0}{v_2} \, (1-2 a_1) \varepsilon^2 \left( \frac{\hbar^2}{m} \, \sum_{i=1}^{N}
|\bm{F}_i|^2  \right) \ . 
\label{doblev}
\end{eqnarray}
The parameters  in Eqs. (\ref{chinaction}) and (\ref{doblev1},\ref{doblev}) are not all
independent and can be written as a function of only two, $a_1$ and $t_0$,
which are restricted to fulfill the conditions 
\begin{eqnarray}
0 & \leq & a_1 \leq 1 \label{a1t01} \\
0 & \leq & t_0 \leq \frac{1}{2} \, \left( 1- \frac{1}{\sqrt{3}} \right) \ .
\label{a1t0}
\end{eqnarray}
The rest of parameters are obtained from the two independent ones
$\{a_1,t_0\}$ according to the equations 
\begin{eqnarray}
u_0 & = & \frac{1}{12} \, \left[ 1 - \frac{1}{1-2t_0} + \frac{1}{6 \,(1-2
t_0)^3} \right] \\
v_1 & = & \frac{1}{6 \, (1-2 t_0)^2}  \\
v_2 & = & 1 -2 v_1 \\
t_1 & = & \frac{1}{2} -t_0  \ .
\label{parameters+}
\end{eqnarray}
The accuracy of the CA depends on the particular values of $a_1$ and $t_0$ that have
to be numerically optimized. Each one modifies the action in different
directions: $t_0$ controls the weight of the different parts in which the
kinetic part is splitted (\ref{chinaction}) and $a_1$ the weight of each  
part in which the double commutator is divided (\ref{doblev}). 

Restricting first our analysis to distinguishable particles, the quantum partition
function $Z$ (\ref{zeta}) can be obtained through a multidimensional
integral of the $M$ terms (beads) in which it is decomposed,
\begin{equation}
Z = \int d \bm{R}_1 \ldots d \bm{R}_M \ \prod_{\alpha=1}^{M} \rho
(\bm{R}_\alpha,\bm{R}_{\alpha+1}) \ ,
\label{zeta2}
\end{equation}           
with $\bm{R} \equiv \{\bm{r}_1,\ldots,\bm{r}_N\}$ and $\bm{R}_{M+1}
=\bm{R}_1$. In the rest of the work, Latin and Greek indexes are used for
particles and beads, respectively. The density matrix of each step in Eq
(\ref{zeta2}) is written in the CA,
\begin{eqnarray}
\lefteqn{\rho (\bm{R}_\alpha,\bm{R}_{\alpha+1}) =}  \nonumber \\
& & \left( \frac{m}{2 \pi \hbar^2 \varepsilon}  \right)^{9N/2} \left(
\frac{1}{2 t_1^2 t_0} \right)^{3N/2} \int d \bm{R}_{\alpha A} \, d \bm{R}_
{\alpha B} \ \exp \left\{ - \frac{m}{2 \hbar^2 \varepsilon}  \right.
\nonumber \\
& & \times \sum_{i=1}^{N} \left( 
\frac{1}{t_1} (\bm{r}_{\alpha,i}-\bm{r}_{\alpha A,i})^2 +  
\frac{1}{t_1} (\bm{r}_{\alpha A,i}-\bm{r}_{\alpha B,i})^2 +
 \frac{1}{2t_0} (\bm{r}_{\alpha B,i}-\bm{r}_{\alpha +1,i})^2 \right) 
 \nonumber  \\
& & - \varepsilon \sum_{i<j}^{N}  \left( v_1 V(r_{\alpha,ij}) + v_2
V(r_{\alpha A,ij}) + v_1 V(r_{\alpha B,ij}) \right)   \nonumber         \\
& & - \left. \varepsilon^3 u_0 \, \frac{\hbar^2}{m} \sum_{i=1}^{N} \left( a_1
|\bm{F}_{\alpha,i}|^2 + (1-2 a_1) |\bm{F}_{\alpha A,i}|^2 + a_1
|\bm{F}_{\alpha B,i}|^2 \right) \right\}  \ .\label{fullrho}
\end{eqnarray}
According to the CA, each elementary block of \textit{width} $\varepsilon$
is splitted into three, with two middle points that we have denoted as $A$ and 
$B$ in the
above expression for the density matrix (\ref{fullrho}). Each one of the
three internal beads resembles a bead in the TIA approximation in the sense
that the beads of the same type $\{ \alpha, \alpha A, \alpha B \}$ 
 interact through a generalized Takahashi-Imada potential, but with different
weights in front of the double-commutator term (\ref{doblev1},\ref{doblev}). The estimators
for the total and partial energies of the system are therefore similar to the ones derived in the TIA
approximation. 

The total and kinetic energy per particle can be readily derived from
first derivatives of the quantum partition function $Z$, 
\begin{eqnarray}
\frac{E}{N} & = & - \frac{1}{NZ} \, \frac{\partial Z}{\partial \beta} 
\label{ener} \\
\frac{K}{N} & = & \frac{m}{N \beta Z} \, \frac{\partial Z}{\partial m} \ ,
\label{kintherm}
\end{eqnarray}
the potential energy being the difference $V/N = E/N - K/N$. The potential
energy can also be derived through the relation~\cite{takahashi}
\begin{equation}
O(\bm{R}) = -  \frac{1}{\beta} \, \frac{1}{Z(V)} \, \left. \frac{d Z(V + \lambda
O)}{d \lambda} \right|_{\lambda=0} \ ,
\label{generaloperator}
\end{equation}
that can be used also for the estimation of other coordinate operators. From the
definition (\ref{kintherm}), which is known as thermodynamic estimator,
the kinetic energy per particle results in
\begin{equation}
\frac{K^{\text th}}{N} = \frac{9}{2 \varepsilon} - \frac{1}{M N} \left(
\frac{m}{2 \hbar^2 \varepsilon^2} \, T_{M N}^{\text t} - \frac{\hbar^2}{m}
\, \varepsilon^2 u_0 W_{M N} \right)  \ ,
\label{kthermo}
\end{equation}
with
\begin{equation}
T_{M N}^{\text t} = \sum_{\alpha=1}^{M} \sum_{i=1}^{N} \left[ \frac{1}{t_1}
\, (\bm{r}_{\alpha,i}-\bm{r}_{\alpha A,i})^2 +  \frac{1}{t_1} \, 
(\bm{r}_{\alpha A,i}-\bm{r}_{\alpha B,i})^2 + \frac{1}{2 t_0} \,
(\bm{r}_{\alpha B,i}-\bm{r}_{\alpha+1,i})^2 \right] \ ,
\label{tmn} 
\end{equation}
and
\begin{equation}
W_{MN} = \sum_{\alpha=1}^{M} \sum_{i=1}^{N} \left[ a_1
|\bm{F}_{\alpha,i}|^2 + (1 -2 a_1) |\bm{F}_{\alpha A,i}|^2 + a_1 
|\bm{F}_{\alpha B,i}|^2 \right]   \ ,
\label{wmn}
\end{equation}

The potential energy can be calculated from the difference between the
total energy and the kinetic one (\ref{kthermo}) or by means of the general
relation (\ref{generaloperator}) with identical result,
\begin{equation}
\frac{V}{N} = \frac{1}{M N} \left( V_{M N} + 2 \, \frac{\hbar^2}{m} \,
\varepsilon^2 u_0 W_{M N} \right) \ ,
\label{potential2}
\end{equation}
with
\begin{equation}
V_{MN} = \sum_{\alpha=1}^{M} \sum_{i<j}^{N}  \left[ v_1 V(r_{\alpha,ij}) +
v_2 V(r_{\alpha A,ij}) + v_1 V(r_{\alpha B,ij}) \right] \ ,
\label{vmn}
\end{equation}
and $W_{MN}$ given by Eq. (\ref{wmn}). This term that appears both in the
total and kinetic energy, but with different weight, comes from the derivative
with respect to $\beta$ and with respect to $m$, respectively, of the  
$W$  potentials (\ref{doblev1},\ref{doblev}), which depend explicitely on the temperature and
the mass.

The variance of the thermodynamic estimation of the kinetic energy
(\ref{kthermo}) can be rather large and increases with the number of beads
$M$.~\cite{herman,janke} This well-known problem is generally solved by  using the virial
estimator~\cite{herman} of the kinetic energy which relies on the invariance of the
partition function under a scaling of the coordinate variables $\bm{r}
\rightarrow \lambda \bm{r}$. The centroid version of the virial estimator
for the CA is given by
\begin{equation}
\frac{K^{\text{cv}}}{N} = \frac{3}{2 \beta} + \frac{1}{M N} \, \left(
\frac{1}{2} \, T_{M N }^{\text v} + \frac{\hbar^2}{m} \, \varepsilon^2 u_0 (W_{M
N} + Y_{M N}) \right) \ .
\label{kinvirial}
\end{equation}  
In this expression, $W_{M N}$ is given by Eq. (\ref{wmn}) and the two other
terms are explicitely
\begin{equation}
T_{M N}^{\text v} =  \sum_{\alpha=1}^{M} \sum_{i=1}^{N}
\left[ v_1 (\bm{r}_{\alpha,i}-\bm{r}_{o,i}) \, \bm{F}_{\alpha,i} +
v_2 (\bm{r}_{\alpha A,i}-\bm{r}_{o,i}) \, \bm{F}_{\alpha A,i} +
v_1 (\bm{r}_{\alpha B,i}-\bm{r}_{o,i}) \, \bm{F}_{\alpha B,i} \right] 
\label{tmnvirial}
\end{equation}
and
\begin{eqnarray}
Y_{M N} & = & \sum_{\alpha=1}^{M} \sum_{i=1}^{N} \sum_{j \neq i}^{N}
\left[ a_1 (r_{\alpha,i} - r_{o,i} )^a \, T(\alpha,i,j)^b_a  \, (F_{\alpha,i} -
F_{\alpha,j})_b  \right.   \nonumber \\
& & + (1-2 a_1) (r_{\alpha A,i} - r_{o,i} )^a \, T(\alpha A,i,j)^b_a \,
(F_{\alpha A,i} - F_{\alpha A,j})_b \nonumber \\
& &  \left. +  a_1 (r_{\alpha B,i} - r_{o,i} )^a \, T(\alpha B,i,j)^b_a \, (F_{\alpha B,i} -
F_{\alpha B,j})_b \right]  \ . \label{ymn}
\end{eqnarray}
In the above expressions, $\bm{r}_{o,i}$ is the center of mass (centroid) of
the chain representing the atom $i$. The indexes $a$, $b$ in the definition
of $Y_{M N}$ (\ref{ymn}) stand for the
Cartesian coordinates and an implicit summation over repeated indices is
assumed. The tensor $T$ appears also in the virial estimation of the
kinetic energy in the TIA approximation and is explicitely given
by~\cite{brualla}
\begin{eqnarray}
T(\gamma,i,j)^b_a & = & \left[ \frac{\delta^b_a}{r_{\gamma,ij}} -
\frac{(r_{\gamma,ij})^b (r_{\gamma,ij})_a}{r_{\gamma,ij}^3} \right] \,
\frac{d V(r_{\gamma,ij})}{d r_{\gamma,ij}}  \nonumber \\
& & + \frac{(r_{\gamma,ij})^b (r_{\gamma,ij})_a}{r_{\gamma,ij}^2}  
\ \frac{d^2 V(r_{\gamma,ij})}{d r^2_{\gamma,ij}}  \ , \label{tensor} 
\end{eqnarray}
where $\gamma$ stands for the three different types of internal beads,
$\alpha$, $\alpha A$, and $\alpha B$, and $\delta^b_a=1$ if $a=b$ and 0
otherwise.

The implementation of the CA in the PIMC algorithm is similar to the one for the TIA.
Going from TIA to CA, one has to split a single bead into three (with different link
lengths), but in these new beads the different atoms  interact
 with a similar generalized potential, 
coming from the double commutator. A simple inspection on the equations of
the density matrix and energies for the CA and TIA shows that the
complexity of both actions are essentially the same and require, for
example, the same order of derivatives of the interatomic potential $V(r)$. 

A final concern that any PIMC calculation must afford is whether the sampling
method has the necessary efficiency in the movement of the chains to 
avoid the slowing down that can appear for long chains when using only
individual bead movements. In this sense, the implementation in the
algorithm of collective smart movements is crucial. To this end, we use the
staging method~\cite{pollock,sprik,tuckerman} combined with movements of the center of mass of each one of
the atoms. In the CA, the length of each chain is not the same and therefore
one has to generalize the staging method. This generalization is discussed
in Appendix A. 

\section{Results}

We have studied the accuracy of the high-order action proposed by Chin in
three different systems: a one-dimensional harmonic oscillator, a drop of
H$_2$ molecules, and bulk liquid $^4$He. The harmonic oscillator is a
very simple system but it has the advantage of its exact analytic solution at
any temperature which allows for an accurate test of the method. The H$_2$
drop is a more exigent test and it has been adopted in previous works as
benchmark for testing different actions. Finally,  liquid $^4$He is
also a well-known system with a wide experimental information available for
comparison and with the appealing feature of remaining liquid even at zero
temperature. In all the simulations we have used a sampling method which
combines movements of the center of mass of the chains and smart movements
of chain pieces with the staging technique.~\cite{pollock,sprik,tuckerman} 
The length of the staging chain and the maximum value of the
displacement of the center of mass are chosen to achieve an acceptance rate
of 30\%-50\%. The results presented below
have been checked to be stable with respect to the frequency and size of
both movements.

\subsection{Harmonic oscillator}

In our first application, we consider a particle in a one-dimensional
harmonic oscillator with Hamiltonian 
\begin{equation}
H=-\frac{\hbar^2}{2m} \frac{\partial^2}{\partial x^2} + \frac{1}{2} m
\omega^2 x^2 \ .
\label{oscil1}
\end{equation}  
As it is well know, this problem can be exactly solved at any temperature
and, in particular, the energy is given by~\cite{kleinert}
\begin{equation}
E=\frac{1}{2} \hbar \omega \coth(\beta \hbar \omega/2) \ .
\label{oscil2}
\end{equation}
In the PIMC simulations we have taken $\omega=\hbar=m=1$ and the
results presented correspond to  temperatures $T=0.1$ and $0.2$. The
energies obtained at $T=0.2$ with
different actions and as a function of the number of beads $M$ are contained in 
Table \ref{t_harmonicoscillator02}, and have to be compared with the exact value 0.50678. 
In order to reproduce the exact energy with these five digits PA requires
the use of $M=512$ and TIA of a quite smaller number $M=32$.~\cite{brualla} This
asymptotic value is reached in the CA case with a sizeable smaller number:
$M=6$ and $M=5$ for $a_1=0$ and $a_1=0.33$, respectively. 

\begin{table}[]
\centering
\begin{ruledtabular}
\begin{tabular}{ccccc}
 $M$  & $E_{\text{PA}}$ &  $E_{\text{TIA}}$ &  $E_{\text{CA}}(a_1=0)$
 &   $E_{\text{CA}}(a_1=0.33)$
  \\
\hline
2   &  0.30755    & 0.44702 &  0.50444 &  0.50643 \\
3   &             &         &  0.50649 &  0.50675  \\
4   &  0.43162    & 0.50053 &  0.50673 & 0.50677  \\
5   &             &         &  0.50677 & 0.50678 \\
6   &             &         &  0.50678 &  0.50678 \\
7   &             &         &  0.50678 &          \\
8   &  0.48424    & 0.50630 &          &   \\
16  &  0.50085    & 0.50675 &          &  \\
32  &  0.50528    & 0.50678 &          &  \\
64  &  0.50641    & 0.50678 &          &  \\
128 &  0.50669    &         &          &     \\
256 &  0.50676    &         &          &  \\
512 &  0.50678    &         &          &   \\
\end{tabular}
\end{ruledtabular}
\caption{PA ($E_{\text{PA}}$), TIA ($E_{\text{TIA}}$), and CA ($E_{\text{CA}}$)
 results for the one-dimensional harmonic oscillator at $T=0.2$. }
\label{t_harmonicoscillator02}
\end{table}

\begin{figure}
\centerline{
        \includegraphics[width=0.7\linewidth,angle=0]{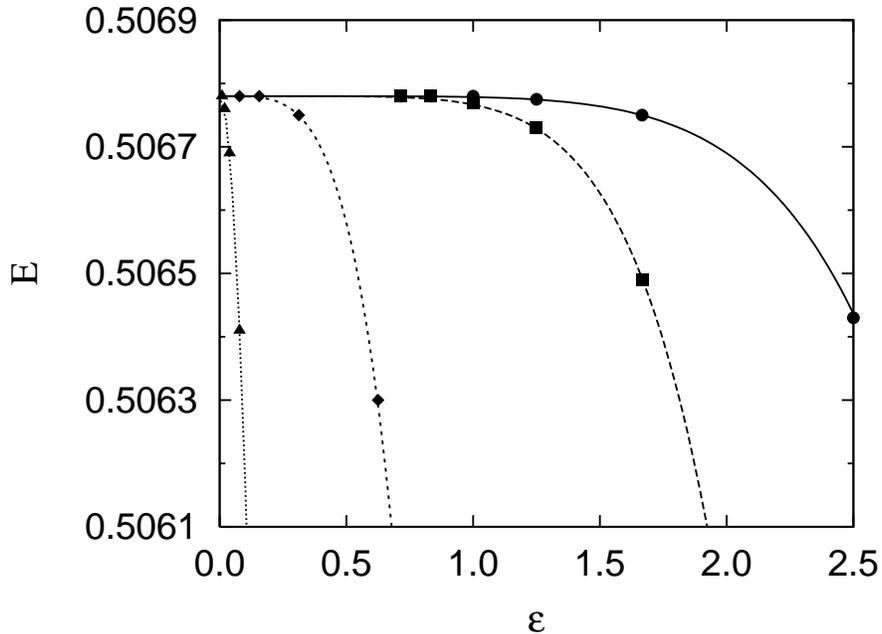}}%
        \caption{PIMC energy of a particle in a one-dimensional harmonic
	well as a function of $\varepsilon$. Triangles, diamonds, squares
	and circles stand for PA, TIA, CA ($a_1=0$),and CA($a_1=0.33$),
	respectively. The lines correspond to polynomial fits
	(\ref{enerfit}) to the data. The errorbars are smaller than the
	size of the symbols.}
\label{f_osci1}
\end{figure}

\begin{figure}
\centerline{
\includegraphics[width=0.7\linewidth,angle=0]{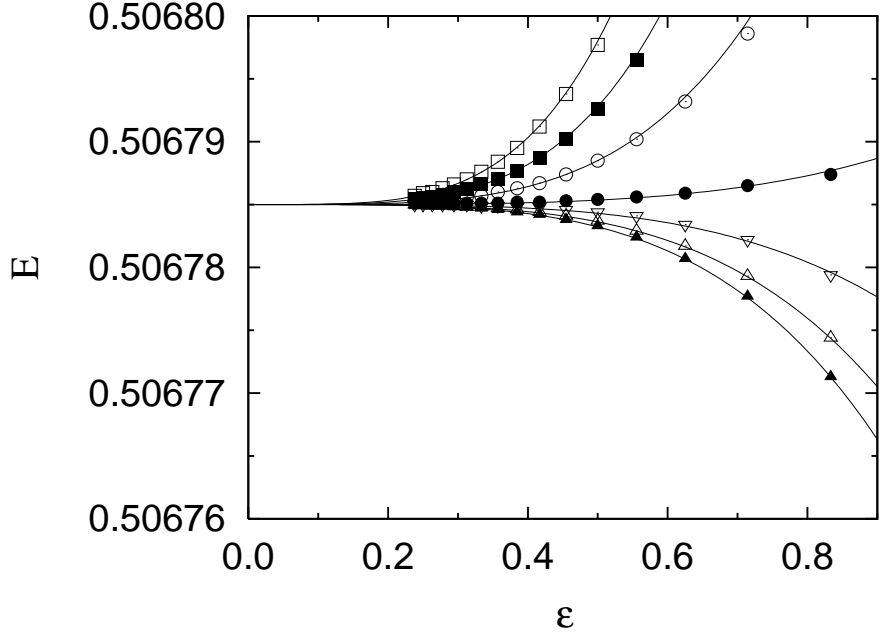}}%
        \caption{Departure from the asymptotic energy $E_0$ for different
	values of $t_0$ and a fixed value for $a_1$ ($a_1=0.33$). From top
	to bottom, $t_0=0.09$, $0.10$, $0.11$, $0.12$, $0.13$, $0.14$, and
	$0.15$. The temperature is $T=0.2$.}
\label{f_osci2}
\end{figure}

In Fig. \ref{f_osci1}, the energies for the different actions at $T=0.2$
are plotted as a function of $\varepsilon$. The lines on top of the PIMC
data correspond to polynomial fits of the form
\begin{equation}
E=E_0 + A_\delta \varepsilon^\delta
\label{enerfit}
\end{equation}
to the energies close to the common asymptotic value $E_0$. From previous
work~\cite{brualla} it is known that $\delta=2$ and 4 for PA and TIA, respectively. In the
case of the CA approximation the departure from $E_0$ is effectively of
sixth order $\delta=6$ in spite of the fact that CA is rigorously a
fourth-order action. This result can be understood as a partial but effective cancellation
between the leading errors of fourth and higher orders due to the fact
that the CA (\ref{chinaction}) is written in terms of two parameters $t_0$ and $a_1$ 
that can be optimized within some constraints(\ref{a1t01},\ref{a1t0}). We show in
Fig. \ref{f_osci2} the characteristic behavior of the energy as a function
of $\varepsilon$ and for different choices of $t_0$ and a fixed value for
$a_1$. The lines are fits to the PIMC data for values of $t_0$ in the range
0.9-0.15 (top to bottom for large $\varepsilon$ values in the figure). When
$t_0 < 0.13$ the asymptotic exact value is approached from above and
contrarily from below when $t_0 \geq 0.13$. By adjusting in a proper way
the value of $t_0$ it is therefore possible to achieve a nearly flat
dependence with $\varepsilon$ and consequently to improve empirically the fourth-order
accuracy of the CA.

\begin{figure}[]
\centerline{
\includegraphics[width=0.7\linewidth,angle=0]{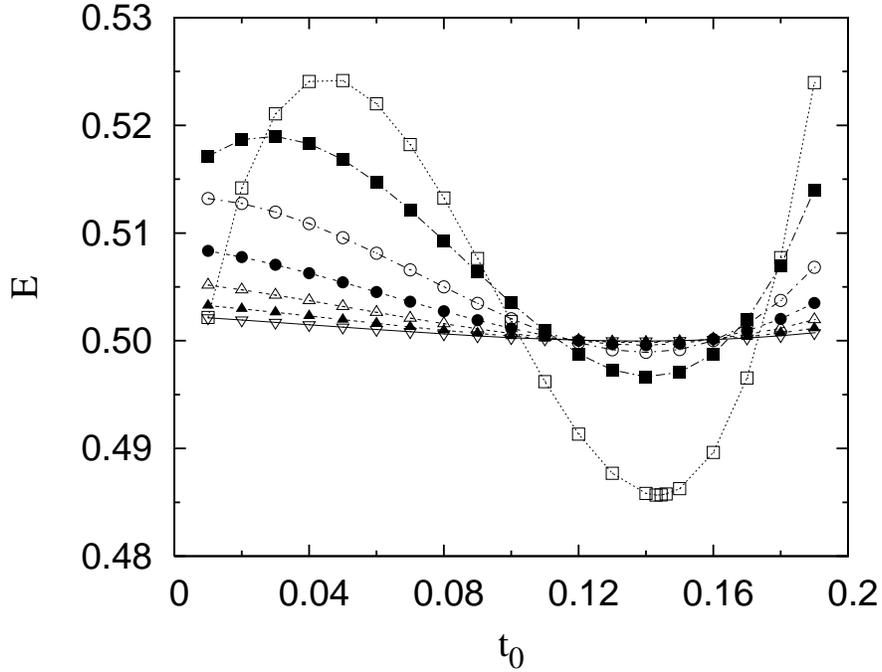}}%
        \caption{Energies as a function of the parameter $t_0$ for
	$a_1=0.33$, and for different number of beads, at $T=0.1$. 
	Squares, $M=2$; filled squares, $M=3$; circles, $M=4$; filled circles,
	$M=5$; up triangles, $M=6$; up filled triangles, $M=7$; down
	triangles, $M=8$. }
\label{f_osci3}
\end{figure}

\begin{figure}[]
\centerline{
\includegraphics[width=0.7\linewidth,angle=0]{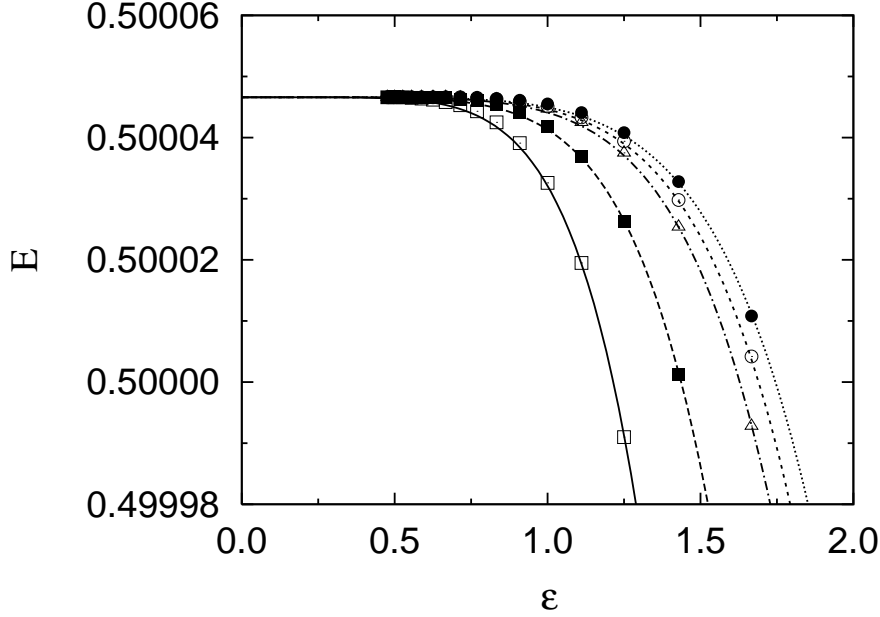}}%
        \caption{Energies as a function of $\varepsilon$ for different
	values of the parameter $a_1$ at $T=0.1$. Squares, $a_1=0$; filled
	squares, $a_1=0.14$; circles, $a_1=0.25$;  filled circles,
	$a_1=0.33$; triangles, $a_1=0.45$. The optimal values of $t_0$ are,
	respectively, $0.1430$, $0.0724$, $0.1094$, $0.1215$, and $0.1298$.
	The lines are polynomial fits (\ref{enerfit}) to the PIMC data.  }
\label{f_osci4}
\end{figure}

The estimation of the optimal value of the parameter $t_0$ for a fixed
$a_1$ can be better made by computing the energies for an increasing number
of beads, starting for a low value. The behavior observed, which is similar for
any allowed $a_1$ value, it is shown in Fig. \ref{f_osci3} for the
particular case of $a_1=0.33$ and $T=0.1$. The lines on top of the data are
guides to the eye and each one of them correspond to a particular $M$ in the
range $M=2$-8. As one can see, there are two small intervals in the $t_0$
scale where the curves tend to intersect (around 0.12 and 0.16) 
and therefore where the
convergence to the asymptotic value is faster. By examining the energies
inside these two regimes one observes that the dependence with $M$ is
slightly smoother in the first one corresponding to smaller $t_0$'s. Thus,
we work in this first regime and normally by selecting a value in which the
approach to the exact energy is from below. For example, for this value
$a_1=0.33$ we have chosen $t_0=0.1215$. It is worth noticing that this
optimal value does not depend on temperature and consequently it can be
adjusted working at higher temperatures where the number of beads required
to achieve convergence is always smaller and thus {\it cheaper} from a
computational point of view. We have verified that the best value of
$t_0$, for a given $a_1$, obtained through this numerical
optimization agrees with the analytical relation for the harmonic oscillator
obtained in Ref. \onlinecite{scuro} ,which
predicts the optimal parameters that cancel exactly the fourth-order error terms.

The optimal $t_0$ depends on the particular value of $a_1$ and  the  
achieved accuracy depends also slightly on $a_1$. This is explicitly shown
in Fig. \ref{f_osci4} where the dependence of the energy on $\varepsilon$
is plotted for values of $a_1$ ranging from 0 to $0.45$. The results
correspond to $T=0.1$ and the lines correspond to fits (\ref{enerfit}) with
$\delta=6$. In all cases the accuracy is of the same order but the best
performance is achieved for $a_1=0.33$ which, according to the expression of
the CA, is the case where the generalized potential $W$
(\ref{doblev1},\ref{doblev}) acts
with the same weight in the three points in which a single step
$\varepsilon$ is splitted.

\subsection{H$_{\bf 2}$ drop}

The case study of a drop composed by a few number of hydrogen molecules has
been used in the past to compare the efficiencies of different PIMC methods
and, in particular, of several approximations for the action. For this
purpose, this system was used for the first time by Chakravarty \textit{et
al.}~\cite{gordillo} to compare the efficiency of Fourier vs. standard PIMC methods. Later
on, it was studied by Predescu \textit{et al.}~\cite{predescu} in a comparative analysis of
energy estimators and by Yamamoto~\cite{yamamoto} in a fourth-order calculation of small
atomic and molecular drops.

\begin{table}[]
\centering
\begin{ruledtabular}
\begin{tabular}{cccc}
 $M$  & $(E/N)_{\text{PA}}$ &  $(E/N)_{\text{TIA}}$ &  $(E/N)_{\text{CA}}$
  \\
\hline  
2   &             & 	         & -40.44(5)  \\
4   &             & 	         & -28.77(3)   \\
8   &  -45.28(3)  & -31.17(3)    & -21.27(2)  \\
10  &             & 	         & 	          \\
12  &             & 	         & -19.13(3)  \\
14  &             & 	         & 	         \\
16  &  -30.59(3)  & -23.48(3)    & -18.32(2) \\
18  &             & 	         & 	          \\
20  &             & 	         & -17.95(3) \\
24  &             & 	         & -17.81(2) \\
28  &             & 	  	 & -17.73(2) \\
32  &  -22.97(3)  & -19.49(3)	 & -17.66(2) \\
36  &             & 	  	 & -17.68(2) \\
48  &    	  &  		 &  -17.68(2) \\
64  &   -19.57(3)  &  -18.05(3)	 &               \\
128 &   -18.28(3)  &  -17.78(3)	 &                \\
256 &   -17.89(3)  &  -17.72(3)	 &              \\
512 &   -17.76(3)  &  		 &               \\
\end{tabular}
\end{ruledtabular}
\caption{PA, TIA, and CA 
 results for the energy per particle of a drop composed by $N=22$ H$_2$ molecules
 at $T=6$ K for different values of $M$. All the energies are in K. The figures in parenthesis are
 the statistical errors affecting the last digit. }
\label{t_hydro}
\end{table}

The drop studied is composed by $N=22$ H$_2$ molecules which are considered
spherical since we restrict our calculation to the $J=0$ state, i.e., to
parahydrogen. The interaction potential is of the form
\begin{equation}
V(\bm{r}_1,\ldots,\bm{r}_N) = \sum_{i<j}^{N} V_2(r_{ij}) + \sum_{i=1}^{N}
V_{\rm c} (r_i)  \ ,
\label{pothydro1}
\end{equation} 
with $V_2$ the intermolecular interaction, assumed to be of Lennard-Jones
type
\begin{equation}
V_2(r_{ij}) = 4 \epsilon \, \left[ \left(\frac{\sigma}{r_{ij}}\right)^{12} - 
\left(\frac{\sigma}{r_{ij}}\right)^6 \right]
\label{ljones}
\end{equation}
and $V_{\rm c}$ is a confining potential introduced to suppress possible
evaporation of molecules,
\begin{equation}
V_{\rm c} (r_i)= \epsilon \left( \frac{|\bm{r}_i -
\bm{R}_{\text{CM}}|}{R_{\rm c}} \right)^{20} \ .
\label{vconfi}
\end{equation}
In $V_{\rm c}$, $\bm{R}_{\text{CM}}=\sum_{i=1}^{N} \bm{r}_i /N$ is the
position of the center of mass of the drop and $R_{\rm c}$ controls the
maximum allowed distance of a particle to  $\bm{R}_{\text{CM}}$. As in
previous work in this problem,~\cite{gordillo} we have chosen $ R_{\rm c}=4 \sigma$, with
Lennard-Jones parameters $\sigma=2.96$ \AA\ and $\epsilon = 34.2$ K
(\ref{ljones}).

\begin{figure}[]
\centerline{
\includegraphics[width=0.7\linewidth,angle=0]{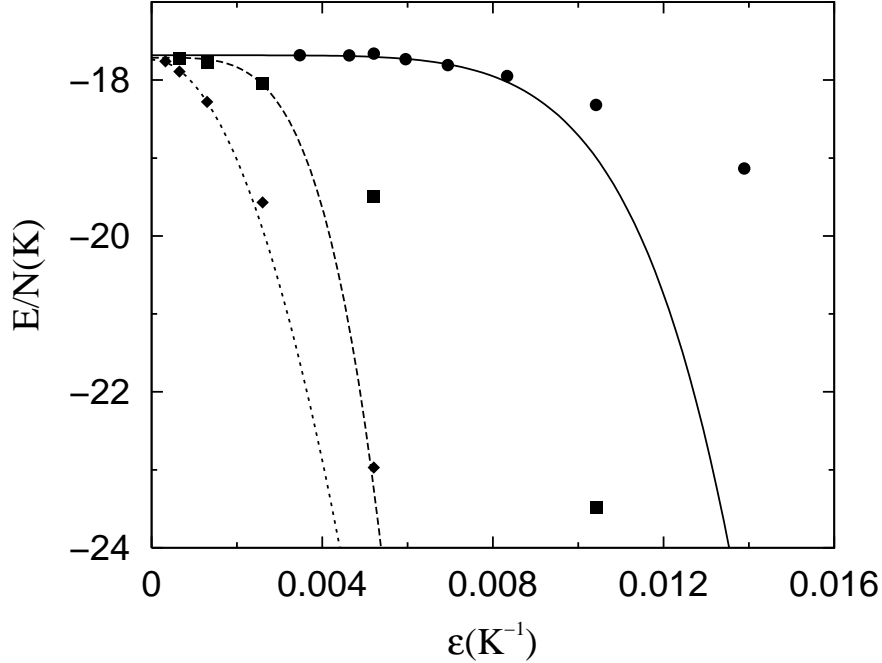}}%
        \caption{Energy of the H$_2$ drop at $T=6$ K as a function of 
	$\varepsilon$ for different actions: PA, diamonds; TIA, squares;
	circles, CA. The parameters of the CA are $a_1=0$ and
	$t_0=0.175$. The lines are polynomial fits (\ref{enerfit}) to the 
	PIMC data.  }
\label{f_hydro1}
\end{figure}

In Table \ref{t_hydro}, we report our results for the energy of the drop
using different number of beads $M$ and several approximations for the
action: PA, TIA, and CA. The asymptotic (unbiased) energy is obtained in
the limit $\varepsilon \rightarrow 0$ ($M \rightarrow \infty$): this is
achieved with $M \simeq 512$, 128, and 32 for PA, TIA, and CA,
respectively. Therefore, also in this more exigent test the CA shows its
appreciably higher efficiency with respect to TIA and other published data
of the same problem obtained with Suzuki actions.~\cite{yamamoto} The value of
$\varepsilon$ required to reach the asymptote in CA is comparable with the one
observed in a previous study of this H$_2$ drop using the pair action
approximation.~\cite{gordillo} As it is shown in Fig. \ref{f_hydro1}, 
the $\varepsilon \rightarrow 0$ behavior is well reproduced
by the polynomial fit (\ref{enerfit}) with different exponents: $\delta=2$,
4, and 6 for PA, TIA, and CA, respectively. Our best result for the energy
of the drop is obtained from the simulation with the CA: the total energy
is $E/N=-17.68(2)$ K, and the potential and kinetic energies are
$V/N=-47.82(3)$ K and $K/N=30.14(2)$ K, respectively. These three energies
are in agreement with the previous estimations of Refs.
\onlinecite{gordillo,predescu,yamamoto}.

\begin{figure}[]
\centerline{
\includegraphics[width=0.7\linewidth,angle=0]{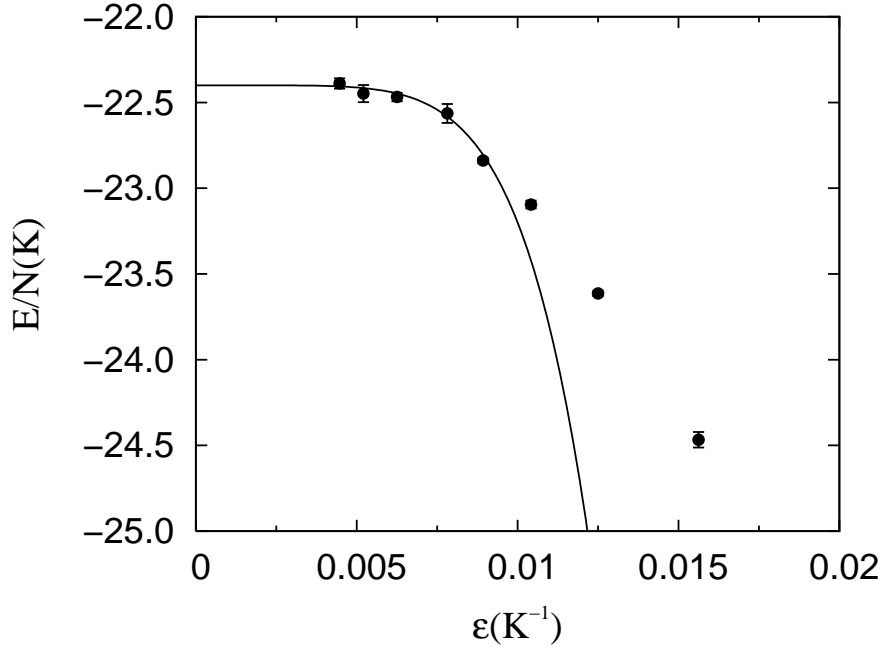}}%
        \caption{Energy of the H$_2$ drop at $T=1$ K as a function of 
	$\varepsilon$ using the CA. The line is  a polynomial fit (\ref{enerfit}) to the 
	PIMC data with $\delta=6$.  }
\label{f_hydro2}
\end{figure}  

Going down in temperature, the PIMC calculation becomes harder due to the
increase in the number of beads necessary to achieve convergence and to the
simultaneous decrease of the acceptance rate in the staging movements. In
order to be effectively able to compute properties in these ultracold
regimes the action in use has to be accurate enough to reduce $M$ to a
manageable level. To this end, we have computed the properties of the
$N=22$ drop at $T=1$ K. In this simulation, as in the one at $T=6$ K, we
have ignored the Bose statistics of the molecules. The dependence of the
total energy with $\varepsilon$ is shown in Fig. \ref{f_hydro2}. As one can
see, the value of $\varepsilon$ ($M=180$) in the asymptote is approximately a factor 
of two smaller
than the one at $T=6$ K (Fig. \ref{f_hydro1}) but a sixth-order behavior is
again observed (solid line in Fig. \ref{f_hydro2}). The energies in the
asymptotic regime are $E/N=-22.47(5)$ K, $V/N=-52.7(2)$ K,and
$K/N=30.2(2)$ K. It is worth noticing
that at this lower temperature PA would require the use of $M \sim 3000$
and TIA of $M \sim 1200$, values which make quite unreliable their use in
the deep quantum regime.

\subsection{Liquid $^{\bf 4}$He}

\begin{table}[]
\centering
\begin{ruledtabular}
\begin{tabular}{ccccc}
 $M$  & $(E/N)_{\text{PA}}$ &  $(E/N)_{\text{TIA}}$ &  $(E/N)_{\text{CA}}$
 ($a_1=0$)  &  $(E/N)_{\text{CA}}$ ($a_1=0.33$)
  \\
\hline  
3   &           &    	     &  -5.59(3)  &   -5.48(3) \\
4   &           &    	     &  -4.51(3)  &   -4.40(3)\\
5   &           &    	     &  -3.91(3)  &   -3.77(3)\\
6   &           &    	     &  -3.53(3)  &   -3.43(3) \\
8   &  -9.29(3) &   -6.30(3) &  -3.16(3)  &   -3.08(3) \\
10  &           &            &  -3.01(3)  &   -2.89(3) \\
12  &           &            &  -2.91(3)  &   -2.83(3)\\
14  &           &      	     &  -2.86(3)  &   -2.81(3)  \\
20  &           &   	     &  -2.81(3)  &           \\
16  &  -5.85(3) &   -3.97(3) &   	  &          \\
32  &  -3.98(3) &   -3.04(3) &   	  &          \\
44  &           &   -2.93(4) &   	  &           \\
64  &  -3.19(4) &   -2.85(4) &   	  &           \\
100 &           &   -2.83(4) &   	  &           \\
128 &  -2.92(4) &   -2.82(4) &   	  &           \\
256 &  -2.84(4) &            &   	  &           \\
512 &  -2.81(4) &      	     &   	  &           \\
\end{tabular}
\end{ruledtabular}
\caption{PA, TIA, and CA ($a_1=0$ and $a_1=0.33$)
 results for the energy per particle of bulk liquid $^4$He 
 at $T=5.1$ K for different values of $M$. All the energies are in K. The figures in 
 parenthesis are
 the statistical errors affecting the last digit. }
\label{t_helium}
\end{table}

As in our previous work,~\cite{brualla} we have studied the accuracy of the CA
in a fully many-body calculation, deep in the quantum regime, as it is liquid
$^4$He. We consider a bulk system at a density $\rho=0.02186$ \AA$^{-3}$
and at two temperatures, $T=5.1$ and $0.8$ K. The calculation is performed
within a simulation box of 64 atoms with periodic boundary conditions and
with an accurate Aziz potential.~\cite{aziz} In order to correct for the finite size of
the system we have added standard potential energy tail corrections relying
on the assumption that beyond $L/2$, with $L$ the size of the box, the
medium is homogeneous, i.e. $g(r) \simeq 1$.

\begin{figure}[]
\centerline{
\includegraphics[width=0.7\linewidth,angle=0]{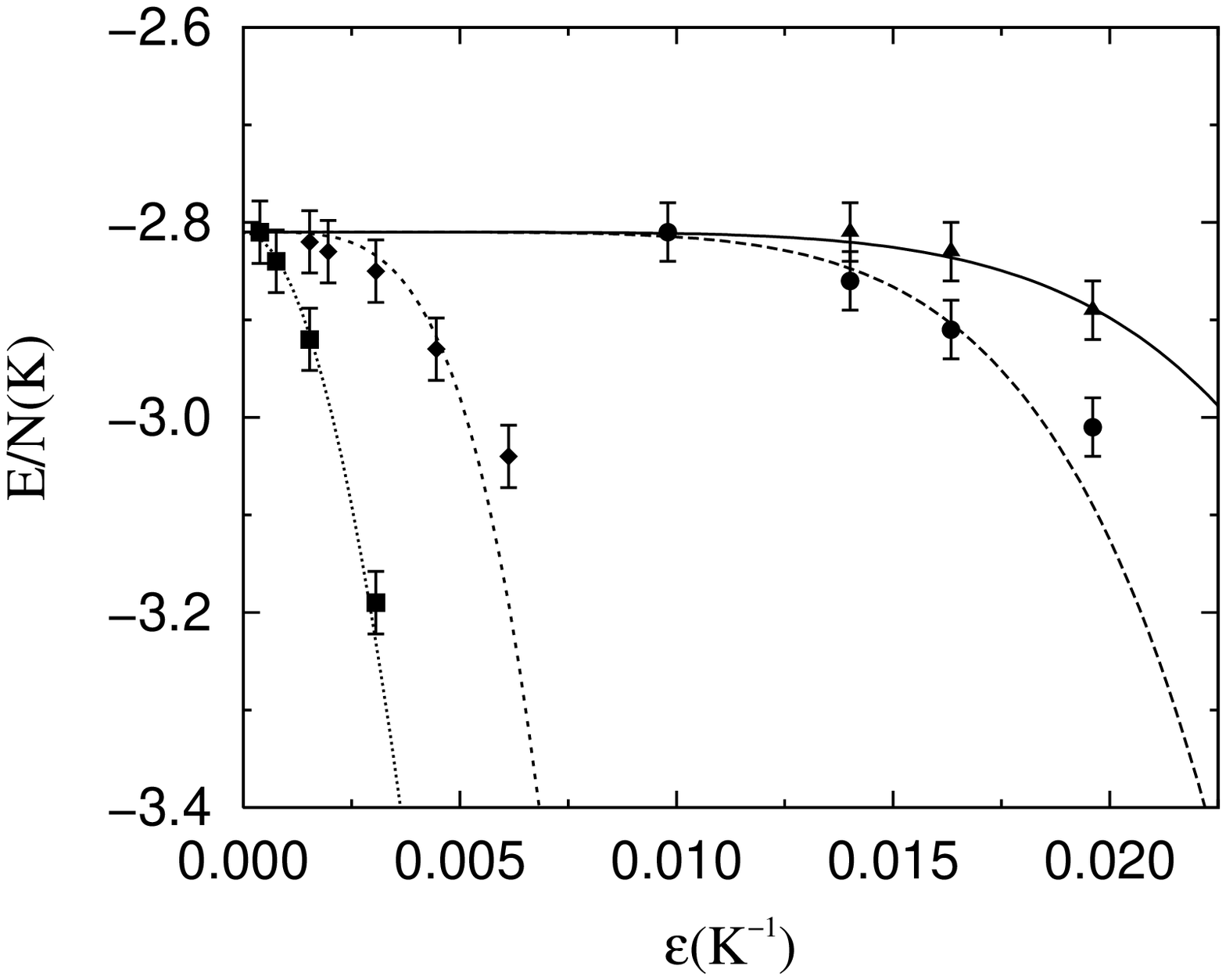}}%
        \caption{Energy per particle of liquid $^4$He at $T=5.1$ K and
	density $\rho=0.02186$ \AA$^{-3}$ as a function of 
	$\varepsilon$ using several actions: PA, squares; TIA, diamonds;
	CA ($a_1=0$), circles; CA ($a_1=0.33$), triangles.  
	 The lines are  polynomial fits (\ref{enerfit}) to the 
	PIMC data with $\delta=2$, 4, and 6 for PA, TIA, and CA,
	respectively.  }
\label{f_helium1}
\end{figure}  

Table \ref{t_helium} contains PIMC results for the energy at $T=5.1$ K using
different number of beads and for several actions: PA, TIA, and CA with
$a_1=0$ and $a_1=0.33$. The parameter $t_0$ has been optimized for both
values of $a_1$: $t_0=0.17$ and $0.082$ for $a_1=0$ and $0.33$,
respectively. As we noted in our previous study,~\cite{brualla} the introduction of the
double commutator within the TIA reduces sizably the number of beads with
respect to PA: the value $M=256$ of PA turns to $M=64$ for the TIA.
The latter $M$ is again considerably reduced by using the CA since the convergence
to the exact energy is reached for a  value as low as $M=12$ ($a_1=0.33$).
The different dependence on $\varepsilon$ of the three models for the action
is shown in Fig. \ref{f_helium1} at $T=5.1$ K. We can observe that the departure
from the asymptote $E_0$ follows the same power-law dependence
(\ref{enerfit}) than in the harmonic
oscillator and H$_2$ drop previously analyzed: $\delta=2$, 4, and 6  
for PA, TIA, and CA, respectively.

\begin{figure}[]
\centerline{
\includegraphics[width=0.7\linewidth,angle=0]{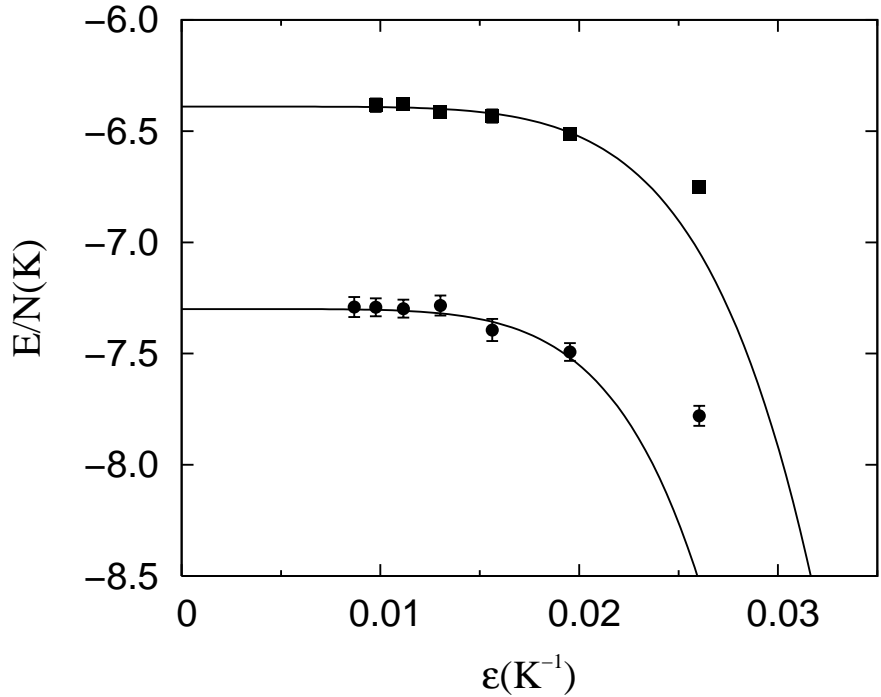}}%
        \caption{Energy per particle of liquid $^4$He at $T=0.8$ K and
	density $\rho=0.02186$ \AA$^{-3}$ as a function of 
	$\varepsilon$ using the CA. Circles and squares correspond to a
	simulation including symmetrization of the density matrix or not,
	respectively.  
	 The lines are  polynomial fits (\ref{enerfit})  with $\delta=6$.}
\label{f_helium2}
\end{figure}  

The accuracy of the CA has been a bit more stressed by repeating the PIMC
simulation at a lower temperature, $T=0.8$ K. At this temperature, the quantum
effects are  bigger than at $T=5.1$ K and the use of TIA, and even more
PA, is completely unreliable due to the large number of beads that are
necessary to eliminate the bias due to a finite $M$ value. At $0.8$ K,
liquid $^4$He  is below the $\lambda$ transition ($T_\lambda=2.17$ K) and
therefore it is  superfluid making absolutely necessary the  
sampling of permutations to accomplish with its boson statistics. The
exchange frequency is drastically reduced over $T_\lambda$ and thus the
inclusion of permutations at the higher temperature ($5.1$ K) does not
modify the results presented above. Results for the energy per particle at
$T=0.8$ K as a function of $\varepsilon$ and using the CA are shown in Fig.
\ref{f_helium2}. The simulations have been carried out including or not the
right symmetry of the thermal density matrix; as one can see, and is well
known, the inclusion of permutations leads to a decrease of the energy which 
is in agreement with the experimental data on this
system. The sampling of permutations is performed within the widely used
method proposed by Pollock and Ceperley.~\cite{ceperley,pollock}  The accuracy of the CA is the same
including or not permutations in the sampling , i.e., sixth order in
$\varepsilon$ (solid lines in Fig. \ref{f_helium2}). The values of
$\varepsilon$ at which the asymptotic trend is observed are similar to the
ones achieved using the pair action,~\cite{ceperley} which is the most widely used
approximation to deal with superfluids within the PIMC formalism.

\section{Conclusions}

\begin{table}[]
\centering
\begin{ruledtabular}
\begin{tabular}{lccc}
     &  CPU cost per bead      &  Decrease of $M$  &   Efficiency  
      \\  \hline
 PA           &     1.0    &    1      &  1.0   \\
 TIA          &     2.9    &    4      &  1.4    \\
 CA           &     7.2    &   58      &  8.0  \\
\end{tabular}
\end{ruledtabular}
\caption{Comparison among the efficiencies of the PA, TIA, and CA in PIMC.  }
\label{t_conclu}
\end{table}

In the last decades there has been a continued effort for improving the
action to be used in PIMC simulations beyond the PA. Working directly on
the exponential of the Hamiltonian, Takahashi and Imada~\cite{takahashi} introduced in the
action the double commutator $[[\hat{V},\hat{K}],\hat{V}]$ and showed that the new algorithm
(TIA) was of fourth order. As showed in previous work,~\cite{brualla} the TIA reduces
significantly the number of beads to reach the asymptotic limit and
therefore it can be very useful in quantum systems if the temperature is
not very small. However, if one is interested on achieving lower
temperatures, deep in the quantum regime, the TIA is still not accurate enough
since the number of beads required is yet too large. As pointed out by
Chin,~\cite{chin} this relative failure of TIA is due to the fact that the TIA is in
fact a fourth-order action but only for the trace. Posterior attempts of
improving the action,~\cite{jang,yamamoto} based on actions proposed by
Suzuki,~\cite{suzuki} did not show a
significant enhancement of the efficiency of the method with respect to
TIA.

Following with the aim of going a step further on the improvement of the
action for PIMC applications, we have used for the first time the new
developments of Chin~\cite{chin,chin2} that have led  to full fourth-order expansions of the
exponential of the Hamiltonian. The resulting action (CA) is more involved
than the TIA and Suzuki action but still it is readily implementable
starting on a TIA approach since the basic ingredients of the CA are
already contained in the TIA. In Table \ref{t_conclu}, we compare the
efficiency of the three models for the action; the numbers correspond to
the calculation of liquid $^4$He but are similar to the ones obtained in
the study of the H$_2$ drop. Considering as 1 the CPU time per bead and the
efficiency in the PA one observes that the cost per bead in TIA is $2.9$
and the number of beads is 4 times smaller, leading to en efficiency
$4/2.9=1.4$. In the CA, the CPU cost per bead is larger since every step
$\varepsilon$ is splitted into three but the big decrease in the number of
beads (58 times smaller than in PA) results in an efficiency 8 times larger
than the PA.   

In the Chin action there are two parameters ($t_0$, $a_1$) that have to be
optimized. The search of these parameters for a given system is not
difficult since they are independent of temperature and therefore they can
be found at higher temperatures where the number of beads is very small and
consequently the simulations very short in CPU time. As we have discussed in
the harmonic oscillator problem, the exact energy is crossed by changing
$t_0$ because the departure from the asymptote can be upwards or downwards
depending on its particular value. This behavior makes possible to search
for optimal parameters that improve the efficiency of the action from
fourth-order to an effective sixth order. This behavior is not only
characteristic of the harmonic oscillator but completely general as we have
verified in real many-body problems as the H$_2$ drop and bulk liquid
$^4$He. This sixth-order law is maintained even at superfluid temperatures
making that the number of beads required for the simulations is completely
manageable. Therefore, the CA is a realistic alternative to the pair action
for dealing with quantum liquids and solids in the superfluid regime at
temperatures close to zero. Its implementation is not much more involved
than the TIA, easier to use than the pair action, and useful also for
problems with non-radial interactions where the application of pair action
is much more involved. Work is also in progress to use the CA in a path
integral ground state (PIGS) approach to study the limit of zero
temperature, the initial simulations showing also a sixth-order accuracy 
and convergence to the exact energies with a few number of beads.

\acknowledgments

We wish to thank stimulating discussions with Siu Chin about high order actions.
Partial financial support from DGI (Spain) Grant No.
FIS2008-04403 and Generalitat de Catalunya Grant No. 2005SGR-00779 are also
aknowledged.

\appendix*
\section{Staging Transformation}

The staging technique allows for a direct sampling of the free (kinetic)
part of the action, i.e., Metropolis test is not necessary. In this
Appendix, we generalize the standard staging method to the one required for
the CA action where the \textit{width} of a given bead depend of its type
($t_1 \varepsilon$, $2 t_0 \varepsilon$) (\ref{fullrho}). If two
points of the chain representing an atom are considered fixed, 
$\bm{r}_0$ and $\bm{r}_M$, one is
interested in transforming the free action between these two extremes 
\begin{equation}
S \equiv \exp \left[ - \sum_{\alpha=1}^{M} c_\alpha (\bm{r}_\alpha -
\bm{r}_{\alpha-1})^2 \right] \ ,
\label{sorig}
\end{equation}     
into the staging one
\begin{equation}
S_{\text{st}} =   C(\bm{r}_0,\bm{r}_M) \, \exp \left\{ -
\sum_{\alpha=1}^{M-1} q_\alpha [ \bm{r}_\alpha -
(a_\alpha \bm{r}_{\alpha-1} + b_\alpha \bm{r}_M) ]^2 \right\} \ .
\label{sstag}
\end{equation}
The constant $C$ depends only on the fixed positions $\bm{r}_0$ and
$\bm{r}_M$ and therefore it is not important for the purpose of the
sampling. By imposing that both actions (\ref{sorig},\ref{sstag}) are equal
one can derive sequential relations for the staging coefficients
$q_\alpha$, $a_\alpha$, and $b_\alpha$ as a function of the known ones of
the original action $c_\alpha$,
\begin{eqnarray}
q_\alpha & =  &c_\alpha + c_{\alpha+1} - q_{\alpha+1} a_{\alpha+1}^2 
\label{qalpha} \\
a_\alpha & = & c_\alpha / q_\alpha  \label{aalpha} \\
b_\alpha & = & ( q_{\alpha+1} a_{\alpha+1} b_{\alpha+1} + c_M
\delta_{\alpha,M-1}) / q_\alpha  \ , \label{balpha}
\end{eqnarray}
with $\delta_{\alpha,M-1}=1$ if $\alpha=M-1$ and 0 otherwise.
These relations have to be applied recursively from $M-1$ to 1 with the
starting conditions $q_M = a_M = b_M = 0$. 

Once all the staging coefficients are determined through Eqs
\ref{qalpha}-\ref{balpha}, the new positions $\bm{r}_\alpha$ are recursively
obtained from 1 to $M-1$ using Gaussian displacements.












\end{document}